\documentstyle[12pt,fleqn]{article}
\topmargin - 0.8cm

\catcode`@=11 \@addtoreset{equation}{section} \catcode`@=12
\mathchardef\SGamma="7100

\begin{document}
\title{\bf Quantum cosmology at the turn of Millennium}
\author{A.O. Barvinsky$^{\dag}$}
\date{}
\maketitle \hspace{-8mm}{\em Theory Department, Lebedev Physics
Institute and Lebedev Research Center in Physics, Leninsky
Prospect 53, Moscow 117924, Russia}
\begin{abstract}
A brief review of the modern state of quantum cosmology is
presented as a theory of quantum initial conditions for
inflationary scenario. The no-boundary and tunneling states of the
Universe are discussed as a possible source of probability peaks
in the distribution of initial data for inflation. It is
emphasized that in the tree-level approximation the existence of
such peaks is in irreconcilable contradiction with the slow roll
regime -- the difficulty that is likely to be solved only on
account of quantum gravitational effects. The low-energy
(typically GUT scale) mechanism of quantum origin of the
inflationary Universe with observationally justified parameters is
presented for closed and open inflation models with a strong
non-minimal coupling.
\end{abstract}
$^{\dag}$Talk given at the IXth Marcel Grossmann Meeting, Rome, July,
2000\\

\section{Introduction}
\hspace{\parindent} This is a generally recognized fact that in
the last two decades the cosmological theory was dominated by the
discovery and rapid progress of the inflation paradigm that scored
explaining the well known paradoxes of the standard big bang
scenario. Interestingly, the beginning of inflation theory was
marked in early eighties by the revival of interest in quantum
cosmology. Before that it was considered merely as a toy model
testing ground for quantum gravity \cite{DW}. However, practically
simultaneously with the invention of the inflation scenario
several suggestions were put forward for special quantum states of
the Universe \cite{HH,H,VilNB,tun,Vil,VilVach} that could serve as
a source of this scenario. Quantum cosmology became the theory of
quantum initial conditions in inflationary Universe.

Such application of this theory was from the start marred by a
number of difficulties of both conceptual and technical nature. To
begin with, a great controversy broke out regarding drastic
difference between two major proposals for the cosmological
wavefunction -- the no-boundary proposal of Hartle and Hawking
\cite{HH,H} (also semiclassically implemented in \cite{VilNB}) and
the tunneling proposal of \cite{tun}. The origin of this
difference is rooted in the peculiarities of quantum gravity
theory (the presence of both positive and negative frequency
solutions of the Wheeler-DeWitt equation, indefiniteness of the
Euclidean gravitational action, problem of time, etc.) which are
not properly understood up till now, so that the fundamentals of
these proposals and the discrepancies of their predictions are
still being disputed \cite{discorde}.

Another problem was that, even within a rather shaky foundation of
the no-boundary and tunneling wavefunctions, in the semiclassical
approximation they would not generate well defined and
sufficiently sharp probability peaks that could serve as a source
of initial conditions for inflation. This is a very general
property of all semiclassical models, and it is very easy to see
this. Indeed, a basic characteristics of the inflation scenario is
an effective Hubble constant $H$, its value determining at later
times all main cosmological parameters of the Universe
    \begin{eqnarray}
    H\Longrightarrow \Omega,\,\rho,\,{\delta T\over T},...\, .
    \end{eqnarray}
In its turn it is usually generated by the inflaton scalar field
$\varphi$, $H=H(\varphi)$, whose initial conditions are determined
by a sharp probability peak in the quantum distribution for
$\phi$. Semiclassically this distribution $\rho(\varphi)$ is given
by the exponentiated Euclidean action of the model $I(\varphi)$ --
the Hamilton-Jacobi function in the classically forbidden domain
       \begin{eqnarray}
       \rho(\varphi)\sim e^{\pm I(\varphi)}.          \label{I1}
       \end{eqnarray}

In the slow-roll regime the inflaton momentum is very small, which
is both true for the Lorentzian and Euclidean domains (classically
allowed and forbidden regions related by analytic continuation).
Therefore the derivative $dI(\varphi)/d\varphi$ is very small, the
graph of $I(\phi)$ is very flat, and it cannot generate any
peak-like behaviour for the corresponding distribution function.
Therefore, slow-roll nature of inflation is always in
irreconcilable contradiction with the demand of the {\em
semiclassical} probability peaks in the wave function.

Finally, in recent years another major objection arose against the
issue of initial conditions for inflation in cosmology, in
general, and in quantum  cosmology, in particular. With the
invention of self-reproducing inflation \cite{inflation,Lindop},
it was understood that the anthropic principle starts playing an
important role \cite{Vilanthr,Guthanthr}. So provided, the
self-reproducing eternal inflation regime is achieved, the total
probability of observing some value of the effective Hubble
constant equals the fundamental quantum probability $P(H)$
weighted by the anthropic probability $P_{\rm anthropic}(H)$ --
the probability of the existence of the observer. The latter is
obviously proportional to the volume of inflating Universe and,
therefore, exponentially depending on time, $P_{\rm
anthropic}(H)\sim \exp(3Ht)$. Therefore, very quickly any
probability peak of $P(H)$ gets wiped out by the anthropic factor
in $P_{\rm total}(H)\sim P(H)\exp(3Ht)$, unless the peak $P(H)$
has a stronger fall off behaviour in $H$ than the exponential one.

In this paper we give a brief overview of the present state of
quantum cosmology and advocate that despite the intrinsic
difficulties and objections of the above type this theory remains
a viable scheme of description for the very early quantum
Universe. In particular, we shall try to show that its imprint on
the observable large scale structure can be used as a testing
ground of fundamental quantum gravity theory in the range of
energies where the semiclassical expansion can be trusted. We
would like to emphasize here that the role of quantum initial
conditions in cosmology should not be underestimated. One of the
reasons is that these conditions determine the energy scale of the
cosmological evolution (encoded in the characteristic value of the
Hubble constant) as compared to both the Planckian scale, where
semiclassical methods break, and the self-reproduction scale of
entering the eternal inflation.

Note that the argumentation against the issue of initial
conditions in cosmology usually starts with the assumption of the
Planckian energy scale at the onset of inflation, resulting in the
self-reproduction conditions for eternal inflation. However, such
a starting point cannot be regarded reliable because of the
absence of non-perturbative methods at high energies and the
absence of fully consistent quantum gravity theory at this energy
scale. Within the conventional perturbation framework the
predictions can be justified only when the entire evolution of the
system stays in the low-energy domain. If this evolution
corresponds to the conventional inflation scenario followed by the
standard big-bang model cosmology evolving down to lower scales,
then the criterion of our semiclassical predictions boils down to
verification of the initial energy scale being much below the
Planckian one. This is the point where the issue of initial
conditions becomes crucially important -- this low initial scale
should not be imposed by hands, but rather be derived from general
assumptions on the cosmological quantum state. One of the goals of
this review is to demonstrate that such a possibility occurs in
the cosmological model with strong non-minimal curvature coupling
of the inflaton -- the model for which the quantum origin of the
Universe turns out to be the low-energy (typically GUT) phenomenon
generating the present day observable cosmological parameters.

The paper is organized as follows. We start with a brief overview
of the state of art in quantum cosmology in the year 2000, discuss
the two fundamental proposals for the cosmological quantum state
-- the no-boundary and tunneling wavefunctions, their implications
in context of the closed and open cosmology and then show how
these states can lead to the low energy phenomenon of the quantum
origin of the inflationary Universe.

\section{Quantum cosmology 2000 -- state of art}
\hspace{\parindent} State of art in present day quantum cosmology
consists of the set of rules, rendering this field a physically
complete and consistent theory, and the scope of approximation
methods that allow one to solve quantum cosmological problems in
concrete setting. Among these main ingredients one can single out
three basic subjects most important for applications: i)
Wheeler-DeWitt equation and the path integral; ii) semiclassical
approximation and beyond; iii) collective variables -- homogeneous
minisuperspace modes and cosmological perturbations.

\subsection{Wheeler-DeWitt equation and path integral}
\hspace{\parindent} The formalism of quantum cosmology is based on
the system of Wheeler-DeWitt equations -- the quantum Dirac
first-class constraints as equations selecting the quantum state
of the gravitational and matter fields
$\Psi[g_{ab}(\textbf{x}),\phi(\textbf{x})]$ in the functional
coordinate representation of canonical commutation relations for
the operators of 3-metric $g_{ab}(\textbf{x})$, matter fields
$\phi(\textbf{x})$ and their conjugated momenta
$p^{ab}(\textbf{x})=\delta/i\delta g_{ab}(\textbf{x})$,
$p(\textbf{x})=\delta/i\delta\phi(\textbf{x})$. These equations
look like
    \begin{eqnarray}
    &&\hat{H}_\bot(\textbf{x})\Psi[g_{ab}(\textbf{x}),
    \varphi(\textbf{x})]=0,
    \nonumber\\
    &&\hat{H}_a(\textbf{x})\Psi[g_{ab}(\textbf{x}),
    \varphi(\textbf{x})]=0.
    \label{1}
    \end{eqnarray}
Here $\hat{H}_\bot(\textbf{x})$ and $\hat{H}_a(\textbf{x})$ are
the operators of the superamiltonian and supermomentum constraints
-- the functional variation operators of the second and first
order correspondingly whose actual form will not be very important
for us.

From the viewpoint of local quantum field theory such operators in
the functional Schrodinger representation are not well-defined
because of poorly defined equal time products of local operators,
but formally this problem is overcome by writing down a formal
path integral solution to these equations -- which all the same
does not control the equal-time commutators
    \begin{eqnarray}
    &&\Psi[g_{ab}(\textbf{x}),\varphi(\textbf{x})]=
    \left.\int
    Dg_{\mu\nu}(x)\,D\phi(x)\,\exp\left({i\over\hbar}
    S[g_{\mu\nu}(x),\phi(x)]\right)
    \right|_{\rm gauge\,\, fixing}, \\
    &&g_{\mu\nu},\phi\Big|_\Sigma
     =g_{ab}(\textbf{x}),
    \varphi(\textbf{x}).                              \label{int}
    \end{eqnarray}
Here the integration runs over 4-metrics and matter fields in
spacetime domain subject to boundary values -- 3-metric
$g_{ab}(\textbf{x})$ and matter field $\varphi(\textbf{x})$ --
arguments of the wavefunction defined on this spacelike boundary.

This integral in the naive form lacking the gauge fixation was
first proposed by H.Leutwyler \cite{Leutwyler} and later derived
in \cite{Bar1,Bar2} with a proper account of the full
Feynman-DeWitt-Faddeev-Popov gauge fixing procedure and boundary
conditions on integration variables (for later and more detailed
formulation see \cite{BarvU,reduc1}). The Euclidean version of
this path integral representation for the solution of
Wheeler-DeWitt equations was then used by Hartle and Hawking in
the formulation of the no-boundary prescription for the
cosmological wavefunction \cite{HH,H}.

\subsection{Semiclassical approximation and beyond}
\hspace{\parindent}Usefulness of the path integral consists, as is
well known, in the possibility of developing a regular
semiclassical expansion. In quantum cosmology, however, the
knowledge of the solution to Wheeler-DeWitt equations -- exact or
within any approximation scheme -- does not guarantee the
exhaustive solution of the physical problem.

One of the reasons is that, in contrast with usual quantum field
theory, the basic Wheeler-DeWitt equation is not evolutionary --
unlike the Schrodinger equation it is hyperbolic rather than
parabolic in its variables, and, moreover, it is not just one
equation, but the whole infinite system of equations, the
consistency of which is guaranteed by the closure of the algebra
of operators in the left-hand side of eqs.(\ref{1}). For this
reason, the formalism of quantum cosmology is devoid of a unique
time variable labeling the evolution, which is nothing but the
manifestation of the diffeomorphism invariance of the theory. This
simple property at the classical level results in disastrous
complications at the quantum level -- sometimes called the problem
of time, the remarkable review of its status given by K.Kuchar
\cite{Karel}. This problem has a number manifestations which,
however, altogether originate from the problem of interpreting the
cosmological wavefunction -- the solution of eqs.(\ref{1}).

In contrast with the evolutionary Schrodinger problem of a
conventional QFT, this interpretation is far from being obvious.
To begin with, the Wheeler-DeWitt equations in view of their
hyperbolic nature imply as a conserved object the quantity which
is not positive definite -- a direct analogue of the situation
with the Klein-Gordon equation in first-quantized theory.
Therefore, this conserved quantity -- inner product in the space
of solutions of the Wheeler-DeWitt equations -- cannot be used for
constructing probability amplitudes. Another side of this problem
is the presence of solutions of both positive and negative
frequencies contributing with opposite signs to the probabilistic
quantities. By and large, this problem has not yet been solved, so
we present here only the existing preliminary steps in its partial
resolution. Interestingly, those steps were undertaken
simultaneously with the construction of semiclassical expansion
for the cosmological wavefunction, which is very often interpreted
as the fact that time (and associated with it conserved
probability) in quantum gravity is not a fundamental concept, but
rather is the notion which arises only in semiclassical
approximation. We believe, though, that such a widespread opinion
is erroneous -- lack of our understanding should not be hidden by
smearing out the fundamental and primary notions of physics.

The first serious attempt to introduce time, probability, etc. in
quantum cosmology originated from the semiclassical approximation
for $\Psi[g_{ab}(\textbf{x}),\phi(\textbf{x})]$ in the sector of
the gravitational field $g_{ab}(\textbf{x})$. This was first done,
at the level of minisuperspace model, by DeWitt in his pioneering
paper on canonical quantum gravity \cite{DW} and then rederived
for generic gravitational system by Rubakov and Lapchinsky in
\cite{RL} and also intensively discussed by Banks \cite{Banks}.
With the wavefunction $\Psi[g_{ab}(\textbf{x}),\phi(\textbf{x})]$
rewritten as
    \begin{eqnarray}
    \Psi[g_{ab}(\textbf{x}),\phi(\textbf{x})]=
    \exp\left({i\over\hbar}S[g_{ab}(\textbf{x})]\right)
    \Psi_m[g_{ab}(\textbf{x}),\phi(\textbf{x})],   \label{II}
    \end{eqnarray}
where $S[g_{ab}(\textbf{x})]$ is the Einstein-Hamilton-Jacobi
function of the pure gravitational field in vacuum, the function
$\Psi_m[g_{ab}(\textbf{x}),\phi(\textbf{x})]$ starts playing the
role of the quantum state of quantized matter fields in the
external classical gravitational background. Such an
interpretation is justified by the fact that, when this function
is restricted to the solution of classical Einstein equations in
vacuum, $g_{ab}(\textbf{x})=g^{\rm class}_{ab}(t,\textbf{x})$,
then it becomes the explicit function of time variable,
$\Psi_m(t)[\phi(\textbf{x})]\equiv \Psi_m[g^{\rm
class}_{ab}(t,\textbf{x}),\phi(\textbf{x})]$. Moreover, on account
of the Wheeler-DeWitt equations it satisfies in the lowest order
approximation in $1/m_P^2$ -- the inverse of the Planck mass
squared -- the Schrodinger equation with the quantum Hamiltonian
of quantized matter fields in {\em external classical
gravitational field without sources}. In operator notations,
$|\Psi_m(t)\big>=\Psi_m(t)[\phi(\textbf{x})]$,
    \begin{eqnarray}
    i\hbar \frac{\partial}{\partial t}\,
    |\Psi_m(t)\big>
    =\int d^3 x\,(N^{\perp}
    \hat {H}_{\perp}+N^a\hat{H}_a)\,
    |\Psi_m(t)\big>,                        \label{III}
    \end{eqnarray}
where the operator Hamiltonian is explicitly written as a linear
combination of matter superhamiltonian and supermomenta with the
background lapse and shift functions as coefficients.

Thus, when the semiclassical approximation for the gravitational
field is reliable, the time variable can be introduced into the
formalism of quantum cosmology by means of the classical
gravitational background that i) neither takes into account the
back reaction of quantized matter nor ii) has its own quantum
fluctuations. The discussion of such a way of introducing time in
cosmology can be found in \cite{RL,BarvU,MukhAnd,BarKief}.
Majority of results in modern cosmology have been obtained within
this interpretation framework -- with time introduced via the
classical background. An obvious limitation of this framework is
that the quantum properties of the gravitational background are
inaccessible and it is incapable of accounting for quantum back
reaction properties -- so, in essence, this is not quantum and
cosmology but rather quantum field theory in curved spacetime. The
origin of this difficulty is obvious -- the quantum effects of the
gravitational and matter fields are not treated on equal footing:
the Shrodinger equation of the above type takes into account
quantum effects of matter exactly (to all powers in $\hbar$), but
disregards all inverse powers of the Planck mass squared
$1/m_P^2$. The way around this difficulty is to develop a regular
semiclassical loop expansion in $\hbar$ without distinguishing
those powers of Planck's constant that arise from $1/m_P^2$ and
those from the quantum loops of matter fields, and, simultaneously
not to loose time and probability interpretation inherent in the
Schrodinger equation above.

This program was implemented in the series of papers
\cite{GenSem,BKr,BarvU,geom} in the lowest non-trivial order of
$\hbar$-expansion for quantum states having the form of a single
semiclassical wave packet
    \begin{eqnarray}
    \Psi[g_{ab}(\textbf{x}),\varphi(\textbf{x})]=
    P[g_{ab}(\textbf{x}),\varphi(\textbf{x})]
    \exp\left({i\over\hbar}
    S[g_{ab}(\textbf{x}),\varphi(\textbf{x})]\right).
    \label{2}
    \end{eqnarray}
Here $S[g_{ab}(\textbf{x}),\varphi(\textbf{x})]$ is the
Hamilton-Jacobi system of the full system of interacting
gravitational and matter fields, and
    \begin{eqnarray}
    P[g_{ab}(\textbf{x}),\varphi(\textbf{x})]=
    P_{\rm 1-loop}[g_{ab}(\textbf{x}),\varphi(\textbf{x})]
    +O(\hbar)
    \end{eqnarray}
is the preexponential factor whose expansion in $\hbar$ begins
with the one-loop term.

In \cite{GenSem,BKr,BarvU} it was shown that the one-loop
preexponential factor can be universally obtained in terms of the
Hamilton-Jacobi function of the system. This factor was obtained
by both solving the Wheeler-DeWitt equations in the approximation
linear in $\hbar$ \cite{GenSem,BKr} and calculating the path
integral in the gaussian approximation \cite{reduc1}. Both methods
give the same result, thus confirming formal consistency of the
theory. The expression for the conserved inner product of
semiclassical quantum states was derived \cite{GenSem,BKr,geom}
(with a nontrivial measure in superspace of 3-metrics and matter
fields). This inner product turned out to coincide with the usual
inner product of states in the physical sector of the theory
arising as a result of the Hamiltonian (or ADM) reduction to true
physical degrees of freedom. Physical time arising in this
reduction was shown to survive the transition to the one-loop
approximation, which means that concept of time is not entirely
semiclassical, but admits continuation to quantum domain. All
these conclusions attained in the one-loop approximation can be
generalized to higher orders of semiclassical expansion by the
price of tecnically complicated formalism -- the only principal
limitation of this framework is that the starting point remains
the semiclassical wave packet of the form (\ref{2}) not mixing
opposite frequencies. This is a fundamental limitation that
reflects conceptual problems in quantum cosmology, that are not
yet resolved.

\subsection{Collective variables -- minisuperspace and cosmological
perturbations} \hspace{\parindent} The success of applications in
a physical theory to an essential extent depends on a particular
approximation scheme used. Together with general semiclassical
expansion, considered above, one should use approximation schemes
that simplify the configuration space of the theory leaving aside
those degrees of freedom which are not very important in the
problem in question. The extremal manifestation of this approach
in cosmology consists in the so called minisuperspace reduction,
when only a finite number of collective degrees of freedom is left
-- describing the spatially homogeneous cosmological model. At the
quantum level, such approximation is not completely consistent,
because the zero-point fluctuations of the discarded degrees of
freedom cannot be excluded by hands -- they might give an
important contribution to the dynamics of those collective
variables that determine the main features of the model. Thus,
more fruitful is the approach when all the degrees of freedom are
retained, but a finite number of them are treated exactly, while
the rest are considered perturbatively in the linearized
approximation and higher. Such a scheme matches well with the
semiclassical expansion in which quantum fluctuations of
linearized modes give contributions to perturbative loop effects.

In application to cosmology, this general scheme implies a well
known theory of cosmological perturbations (see, for example
\cite{BMF}), when the metric and matter field are decomposed into
the spatially homogeneous background and inhomogeneous
perturbations
        \begin{eqnarray}
        &&ds^2=-N^2(t)dt^2+a^2(t)\gamma_{ij}dx^i dx^j
        +h_{\mu\nu}(x)dx^\mu dx^\nu,                   \label{2.7a}\\
        &&\varphi(x)=\varphi(t)+\delta\varphi(x), \,\,\,
        x=(t,\mbox{\bf x}),                            \label{2.7}
        \end{eqnarray}
where $a(t)$ is the scale factor, $N(t)$ is the lapse function and
$\gamma_{ij}$ is the homogeneous spatial metric (for closed
cosmology this is a metric of the 3-sphere of unit radius). The
full set of fields consists of the minisuperspace sector of
spatially homogeneous variables $(a(t),\,\varphi(t),\,N(t))$ and
inhomogeneous fields $f(x)$ essentially depending on spatial
coordinates $x^i$={\bf x}
        \begin{eqnarray}
        f(x)=\delta\varphi(t,\mbox{\bf x}),
        \,h_{\mu\nu}(t,\mbox{\bf x}),
        \,\chi(t,\mbox{\bf x}),\,\psi(t,\mbox{\bf x}),\,
        A_\mu(t,\mbox{\bf x}),...                      \label{2.8}
        \end{eqnarray}
The role of classically non-vanishing scalar field $\varphi$ (or
the field with non-vanishing expectation value) is usually played
by inflaton that drives the quasi-exponential expansion at the
inflationary stage of the evolution. On the other hand, the
cosmological perturbations of metric, the inflaton field itself
and other matter fields  -- initially quantum and later
semiclassically coherent -- describe the formation of the large
scale structure on the Robertson-Walker cosmological background
(including microwave background radiation and other matter in the
observable Universe).

The description of cosmological perturbations requires the
formalism that deals with its gauge invariance properties
\cite{Bardeen,BMF}-- not all the variables above are dynamically
independent and physically significant, because part of the
variables represent just purely coordinate degrees of freedom or
those degrees of freedom that can be excluded in virtue of
constraints in terms of physical variables. The dynamical content
of the latter is basically the following. The spatially
homogeneous sector of inflaton $\varphi$, scale factor $a$ and
lapse $N$ gives rise to only one dynamical degree of freedom -- it
can be without loss of generality identified with the inflaton
$\varphi$, while for the inhomogeneous perturbations, basically it
is transverse, transverse-traceless, etc., components that form
the physical sector. We shall collectively denote them by $f^T$.

With such a decomposition, the effective dynamics of the main
collective variable in cosmology -- the inflaton field $\varphi$
is determined by the reduced density matrix obtained by tracing
out the inhomogeneous fields $f^T$. If in the physical sector the
full cosmological state is denoted by
$\mbox{\boldmath$\Psi$}(\varphi,f^T)$, then this density matrix is
given by
        \begin{eqnarray}
        \hat\rho\equiv\rho(\varphi,\varphi')=\int df^T
        \mbox{\boldmath$\Psi$}(\varphi,f^T)
        \mbox{\boldmath$\Psi$}^*(\varphi',f^T).         \label{2.25}
        \end{eqnarray}

The diagonal element of this density matrix is the distribution
function of the inflaton $\varphi$,
        \begin{eqnarray}
        \rho(\varphi)=\rho(\varphi,\varphi),         \label{2.26}
        \end{eqnarray}
which might yield initial conditions for inflation, provided it
has a peak-like behaviour for some suitable mean values of
$\varphi$. This goes as follows.

In the chaotic inflation model the stage of inflation is generated
within the slow roll approximation, when the inflaton field slowly
rolls down the the potential $V(\varphi)$ -- some monotonically
growing function of $\varphi$ -- which, in its turn, determines a
big value of the effective Hubble constant, $\dot{a}/a\simeq
H(\phi)$,
        \begin{eqnarray}
        H^2(\phi)=\frac{8\pi V(\varphi)}{3m_P^2}.   \label{3.14}
        \end{eqnarray}
The initial value of $\varphi$ actually determines all main
cosmological parameters, including the duration of inflation in
terms of the e-folding number $N$ -- the logarithmic expansion
coefficient for the cosmological scale factor $a$ during the
inflation stage,
        \begin{eqnarray}
        N=\int_0^{t_F}dt H                     \label{i2}
        \end{eqnarray}
(with $t=0$ and $t_F$ denoting the beginning and the end of
inflation epoch) and  the present day value of $\Omega$
\cite{HawkTur},
        \begin{eqnarray}
        \Omega\simeq\frac1{1\mp B\exp(-2N)},        \label{i1}
        \end{eqnarray}
The signs $\mp$ here are related respectively to the closed and
open models, and $B$ is the parameter incorporating the details of
the reheating and radiation-to-matter transitions in the early
Universe. Depending on the model for these transitions, its order
of magnitude can vary from $10^{25}$ to $10^{50}$ (when the
reheating temperature varies from the electroweak to GUT scale).
In what follows we shall assume the latter as the most probable
value of this parameter.

Eq. (\ref{i1}) clearly demonstrates rather stringent bounds on
$N$. For the closed model  the e-folding number should satisfy the
lower bound $N\geq\ln B/2\simeq 60$ in order to generate the
observable Universe of its present size, while for the open model
$N$ should be very close to this bound $N\simeq 60$ in order to
have the present value of $\Omega$ {\em not very close to zero or
one}, $0<\Omega<1$, the fact intensively discussed on the ground
of the recent observational data.

In the chaotic inflation model the effective Hubble constant is
generated by the potential of the inflaton scalar field and all
the parameters of the inflationary epoch, including its duration
in units of $N$, can be found as functions of the initial value of
the inflaton field $\varphi$ at the onset of inflation $t=0$. If
this initial condition belongs to the quantum domain then it has
to be considered subject to the quantum distribution
(\ref{2.25})-(\ref{2.26}) following from the cosmological
wavefunction. If this distribution function has a sharp
probability peak at certain $\varphi$, then, at least within the
semiclassical expansion, this value of $\varphi$ serves as the
initial condition for the inflationary dynamics.

\section{No-boundary and tunneling wave functions}
\hspace{\parindent} Quantum states that give initial conditions
for inflation were suggested in early eighties. Basically, these
are two states having in the semiclassical approximation
qualitatively different behaviours. One of these states -- the so
called no-boundary one -- was suggested by Hartle and Hawking
\cite{HH,H} in the form of the Euclidean path integral
prescription. This prescription reduces to the Euclidean version
of the integral (\ref{int}) where the integration goes over
Euclidean compact 4-geometries and 4-dimensional histories of
matter fields "bounded" by the 3-geometry and matter field -- the
arguments of the cosmological wavefunction. The underlying
spacetime of Euclidean signature has the topology of a
4-dimensional ball. Another quantum state -- the tunneling one --
was suggested as a particular semiclassical solution of the
minisuperspace Wheeler-DeWitt equation in the Robertson-Walker
model with the inflaton potential, generating the effective Hubble
(or cosmological constant) \cite{tun,Vil,VilVach}.

The both states were analyzed in much detail for the model of
minimally coupled inflaton field $\varphi$ having a generic
potential $V(\varphi)$
        \begin{equation}
        S[g_{\mu\nu},\varphi]=\int d^{4}x\, g^{1/2}
        \left(\frac{m_P^2}{16\pi}\,R(g_{\mu\nu})-
        \frac12(\nabla\varphi)^2
        -V(\varphi)\right).         \label{min}
        \end{equation}
As semiclassical solutions of the minisuperspace Wheeler-DeWitt
equation in this model, they can be written down in the slow roll
approximation (when the derivatives with respect to $\varphi$ are
much smaller than the derivatives with respect to $a$). They read
\cite{Vil,VilVach}
        \begin{eqnarray}
        &&\mbox{\boldmath$\Psi$}_{NB}(\varphi,a)=
        C_{NB}(a^2H^2(\varphi)-1)^{-1/4}
        \exp\left[-\frac12 I(\varphi)\right]
        \cos\left[S(a,\varphi)+\frac\pi4\right],      \label{3.11} \\
        &&\mbox{\boldmath$\Psi$}_{T}(\varphi,a)=
        C_{T}(a^2H^2(\varphi)-1)^{-1/4}
        \exp\left[+\frac12 I(\varphi)
        +iS(a,\varphi)+\frac{i\pi}4\right]              \label{3.12}
        \end{eqnarray}
and describe two types of the nucleation of the Lorentzian
quasi-DeSitter spacetime (described by the Hamilton-Jacobi
function $S(\varphi,a)$) from the gravitational semi-instanton --
the Euclidean signature hemisphere bearing the Euclidean
gravitational action $I(\varphi)/2$
        \begin{eqnarray}
        &&I(\varphi)=
        -\frac{\pi m_P^2}{H^2(\varphi)},    \nonumber \\
        &&S(\varphi,a)=
        -\frac{\pi m_P^2}{2H^2(\varphi)}
        (a^2H^2(\varphi)-1)^{3/2}.                   \label{3.13}
        \end{eqnarray}
The size of this hemisphere -- its inverse radius -- as well as
the curvature of the quasi-DeSitter spacetime are determined by
the effective Hubble constant (\ref{3.14})

In the tree-level approximation the quantum distributions of
universes with different values of the inflaton field $\phi$
(\ref{2.25})-(\ref{2.26}) are, thus, given by two algorithms --
the real amplitudes of (\ref{3.11})-(\ref{3.12}):
        \begin{eqnarray}
        \rho_{\rm NB}(\varphi)
        \sim e^{-I(\varphi)}          \label{i0}
        \end{eqnarray}
for the no-boundary quantum state \cite{HH,H,VilNB} and
        \begin{eqnarray}
        \rho_{\rm T}(\phi)\sim e^{\,I(\varphi)},     \label{i00}
        \end{eqnarray}
for the tunneling one \cite{tun}. For the {\it minimally} coupled
inflaton, $I(\varphi)$ -- the action of the gravitational
instanton -- reads, up to inflationary slow roll corrections, as
    \begin{eqnarray}
    I(\phi)\simeq -\frac{3m_P^4}{8V(\varphi)},      \label{i000}
    \end{eqnarray}
where $V(\varphi)$ is the inflaton potential. From these equations
it immeadiately follows that the no-boundary and tunneling
wavefunctions lead to opposite predictions: most probable
universes with a minimum of the inflaton potential in the
no-boundary case and with a maximum  -- for the tunneling
situation \cite{Vil}.

However, for reasons discussed in Introduction these extrema in
the probability distribution cannot generate inflationary
scenario. The main obstacle is the irreconcilable contradiction
between the slow-roll nature of inflation and the requirement of
sharp probability peaks -- for slow-roll regime the
$\varphi$-derivatives of the distribution function are very small,
because by order of magnitude they coincide with the momentum
conjugated to $\varphi$ (or rate of change of $\varphi$), and,
therefore, $\rho_{\rm NB,T}(\varphi)$ cannot have sharp enough
peaks in the inflationary domain. The only means possible for
overcoming this difficulty is, to the best of our knowledge, to go
beyond the tree-level approximation. The loop part of the
distribution function, depending on the cosmological model, can
qualitatively change the predictions of the tree-level theory. We
shall show this below for the model with non-minimally coupled
instanton accounting for one-loop effects in the quantum ensemble
of tunneling Universes.

\subsection{Hawking-Turok wavefunction and open inflation}
\hspace{\parindent} Another interesting application of quantum
cosmology was the attempt to generate open inflation from
Euclidean quantum gravity similarly to the case of the closed
Universe. Hawking and Turok have recently suggested the mechanism
of quantum creation of an open Universe from the no-boundary
cosmological state \cite{HawkTur}. Motivated by the observational
evidence for the potential possibility of $\Omega<1$ and the
necessity to avoid a rather contrived nature of inflaton
potentials in the early suggestions for open inflation
\cite{Bucher,Linde} (see also \cite{Linde1}) they constructed a
special gravitational instanton. Within the framework of the
no-boundary cosmological wavefunction this instanton is capable of
generating expanding open homogeneous universes without assuming
any special form for the potential. The prior quantum probability
of such universes weighted by the anthropic probability of galaxy
formation was shown to be peaked at $\Omega\sim 0.01$.

Very briefly, the construction of the Hawking-Turok instanton
generating the open inflation, as compared to the closed one, is
as follows. The inflating Lorentzian spacetime originates in both
closed and open models by the nucleation from a 3-dimensional
section of the gravitational instanton. In the closed model this
is the equatorial section -- the boundary of the 4-dimensional
quasi-hemishpere labelled by the constant value of the latitude
anglular coordinate. The analytic continuation of this coordinate
into the complex plane gives rise to the Lorentzian quasi-DeSitter
spacetime modelling the open inflation. In the open case the
Hawking-Turok suggestion was to continue the Euclidean solution
beyond the equatorial section up to the point where the Euclidean
scale factor again vanishes at the point antipoidal to the regular
pole on the first hemisphere. The nucleation surface then has to
be chosen as the longitudinal section of this quasi-spherical
manifold passing through the regular pole and its antipoidal
point. Then the analytic continuation of the corresponding
longitudinal angle into the complex plane gives rise to the
Lorentzian spacetime. The light cone originating from the regular
pole cuts in this spacetime the domain sliced by the open
spatially homogeneous sections of constant negative curvature.
Their chronological succession serves as the model for the open
inflationary Universe.

This idea, despite its extremely attractive nature, was criticized
from various sides. The Hawking-Turok instanton turned out to be
singular, and its singularity raised a number of objections. They
were based on the possible instability of flat space \cite{Vilen},
originated from the Euclidean theory
\cite{BousLind,Unruh,Wu,Garriga} and from the viewpoint of the
resulting timelike singularity in the expanding Universe
\cite{Unruh,Starobin,Garriga}. The criticism of singular
instantons was followed by attempts of their justification by a
special treatment of spacetime boundary near the singularity
\cite{HawkTur1} and by considering them as a dimensional reduction
from instantons in five dimensions \cite{Garriga1} and in
11-dimensional supergravity \cite{Reall,Breme}. Despite the
pessimistic conclusions of \cite{Reall} on the impossibility of
inflation in supergravity induced model, it is still very likely
that the Hawking-Turok instantons should not be completely ruled
out, because their singularity probes Planckian scales where
classical equations fail and strong quantum effects might regulate
arising infinities. At least, their issue can be regarded open as
long as the Hawking-Turok mechanism promises interesting
predictions.

The latter was, maybe, the main reason of descending interest in
the whole idea, because the practical goal of quantum cosmology --
generating the open Universe with observationally justified modern
value of $\Omega$, not very close to one or zero, -- has not been
reached. Moreover, the open inflation mechanism essentially used
anthropic principle. This, as recognized by the authors of
\cite{HawkTur}, is certainly a retreat in theory, because it makes
one to give up on the goal of explaining all the properties of the
Universe by using some to constrain others. Another difficulty in
the theory of quantum origin of modern Universe arises when it can
undergo the self-reproducing inflation scenario \cite{inflation}.
According to \cite{Lindop}, this scenario washes away in the
observable cosmological data any imprint of quantum initial
conditions and, thus, makes their setting meaningless.

In the discussion of these difficulties Linde suggested to replace
the no-boundary framework of the open inflation by the approach of
tunneling quantum state of the Universe \cite{Lindop}. Such a
replacement, according to the discussion above, shifts the
probability to larger values of the Hubble constant $H(\phi)\sim
V^{1/2}(\phi)$ and eventually increases the amount of inflation
and the value of $\Omega$. Then, provided that there is an upper
bound on the inflaton or its potential beyond which inflation is
impossible, one can get at the point of this bound a particular
value of $\Omega$ fitting into a needed observational range.

This idea was realized by Linde in \cite{Lindop} for two classes
of models -- nonminimally coupled inflaton field with the {\it
positive} coupling constant $\xi$ in the interaction term
$-\xi\varphi^2 R/2$ \cite{GBLinde,Linde} and the model with
supergravity induced exponential potentials \cite{LindeRiotto}. In
these models the boundary of the inflation domain naturally
followed from the positivity of the effective gravitational
constant (overall coefficient of the curvature scalar, see next
section for details), $m_P^2/16\pi-\xi\varphi^2/2>0$, or the
steepness restrictions on the inflaton potential and could
generate a limited amount of inflation with a needed value of
$\Omega$. The latter property could guarantee the absence of
conditions necessary for self-reproducting inflation scenario and,
thus, protect the model from washing out the initial conditions in
this scenario \cite{Lindop}.

However, a simple qualitative analysis of the above type has a
number of limitations. First of all, in the model with nonminimal
coupling the instanton action, even in the lowest order of the
slow roll expansion, is given by a more complicated expression
than (\ref{i000}) (see eq. (\ref{1.2}) below in case of the
quartic potential $\lambda\varphi^4/4$). This becomes especially
important in case of a big {\it negative} nonminimal coupling
constant, $\xi<0$, which is often regarded preferable from the
viewpoint of the CMBR anisotropy \cite{nonmin1,nonmin2}, because
it solves the problem of excessively small $\lambda$ for the
minimal inflaton \cite{inflation} (it allows one to trade a very
small $\lambda$ in favour of a big $|\xi|\gg 1$ in the expression
for $\Delta T/T\sim\sqrt{\lambda}/|\xi|\sim 10^{-5}$ in this
model). Thus, the expression (\ref{i000}) cannot be directly used
in the nonminimal model. Secondly, the slow roll corrections
should be taken into account, and they become particularly
important on the Hawking-Turok instanton in view of its
singularity. Finally, tunneling processes in both models of
\cite{Lindop} (nonminimal and supergravity induced) start at
Planckian energies much beyond reliable perturbative domain, where
conventional semiclassical methods are not applicable. This means,
that at least lowest order loop effects should be considered and
serious arguments found for the justification of loop expansion.
Below we show, that with the inclusion of loop effects the
mechanism of creation of the inflationary Universe is also
possible in the open case, and, as in the closed model, this
mechanism belongs to the low-energy domain, which justifies the
semiclassical methods.

\section{Quantum origin of the Universe as a low-energy phenomenon}
\hspace{\parindent} One can try resolving the difficulties of the
above type by resorting to two possibilities: i) changing the
Lagrangian of the inflationary model and ii) by going beyond the
tree-level approximation. Although the model (\ref{min}) captures
all essential features of chaotic inflation theory, there exists
important low-energy modification (that is the modification that
does not involve curvature-squared and of higher powers in
curvatures terms essential only for Planckian scales) that can
qualitatively change the above conclusions. This is the model with
the inflaton field non-minimally coupled to curvature. Below we
consider this model which, together with the inclusion of quantum
gravitational loop effects, renders the mechanism of the
low-energy phenomenon of the quantum origin of the inflationary
Universe compatible with observations.

\subsection{Non-minimal inflaton coupling} \hspace{\parindent}
Inflation with non-minimally coupled inflaton is described by the
Lagrangian with the inflaton-graviton sector
        \begin{equation}
        {\mbox{\boldmath $L$}}(g_{\mu\nu},\varphi)
        =\frac{m_{P}^{2}}{16\pi} R(g_{\mu\nu})
        -\frac{1}{2}\xi\varphi^{2}R(g_{\mu\nu})
        -\frac{1}{2}(\nabla\varphi)^{2}
        - V(\varphi).        \label{0.2}
        \end{equation}
Such a model has a good family of inflationary solutions for  a
big negative constant of non-minimal coupling constant
$-\xi=|\xi|\gg 1$.

This model is of a particular interest for a number of reasons.
Firstly, from the phenomenological viewpoint a strong nonminimal
coupling allows one to solve the problem of exceedingly small
$\lambda$. Here the observable magnitude of CMBR anisotropy
$\Delta T/T\sim 10^{-5}$ is proportional to the ratio
$\sqrt{\lambda}/|\xi|$ \cite{nonmin1,nonmin2}), so that instead of
exceedingly small value of $\lambda\sim 10^{-13}$ (unacceptable
from the viewpoint of reheating theory) one can take big $|\xi|$
to satisfy the observational constraint. Secondly, this coupling
is inevitable from the viewpoint of renormalization theory. Also,
among recent implications, it might be important in the theory of
an accelerating Universe \cite{BEPStar}.

Another advantage of this model is the fact that, in the slow-roll
regime, it is qualitatively equivalent to the minimal model but
with the effective, $\varphi$-dependent, Planck mass constant --
the overall coefficient of the curvature square
        \begin{eqnarray}
        m_P^2\rightarrow
        m_{\rm eff}^2(\varphi)=m_P^2+8\pi|\xi|\varphi^2.
        \end{eqnarray}
For large $|\xi|\varphi^2$ this constant is much bigger than the
original Planck mass, therefore it strongly improves the
perturbation expansion in graviton loops, because this expansion
runs in inverse powers of $m_{\rm eff}^2(\varphi)$.

Important modification of the distribution function in quantum
cosmology of this model occurs already at the tree-level
approximation. Now the Euclidean instanton action takes the form
        \begin{eqnarray}
        I(\varphi)\simeq-\frac{3m_{\rm eff}^4(\varphi)}
        {8V(\varphi)},                                    \label{Inon}
        \end{eqnarray}
and in view of nontrivial $\varphi$-dependence of $m_{\rm
eff}^4(\varphi)$ the shape of the distribution function can
acquire new interesting probability peaks. Interestingly, for a
particular inflaton potential this model actually suggest a
natural resolution of the problem mentioned in Introduction -- the
wiping out of initial conditions for inflation due to
exponentially growing anthropic factor. Indeed, if one takes the
simplest quadratic potential without higher (quartic) terms, then
as one can check it will correspond to approximately constant,
$\varphi$-independent, value of the Hubble constant,
        \begin{eqnarray}
        V(\varphi)={1\over2}m^2\varphi^2,\,\,\,
        H^2\sim\frac{m^2}{6|\xi|}={\rm const},
        \end{eqnarray}
and give rise to the total probability -- the fundamental {\em
tunneling} probability factor times the anthropic factor
        \begin{eqnarray}
        \rho^{\rm total}_T(\varphi)\sim
        \exp\left(-\frac{3m_P^4}{4m^2\varphi^2}
        -\frac{48\pi^2\xi^2\varphi^2}
        {m^2}+\frac{3mt}{\sqrt{6|\xi|}}\right).
        \end{eqnarray}
This distribution has a strong damping for both large and small
$\varphi$, while the anthropic factor at all turns out to be
$\varphi$-independent. Therefore, the probability peak at
$\varphi\sim m_P/\sqrt{|\xi|}$ never gets washed away due to
anthropic aspects of self-reproducing inflation.

Unfortunately this theory can hardly be considered seriously,
because the restriction by simplest quadratic potential -- the
crucial property guaranteeing the above quasi-gaussian in
$\varphi$ behaviour -- is not consistent from the viewpoint of
quantum theory. Renormalization effects would always generate
quartic and other terms in the inflaton potential which would
immediately destroy this nice picture. So let us consider another
source of possible mechanism for initial conditions for inflation
-- quantum loop contributions to the distribution function.

\subsection{Beyond tree level -- one-loop effect of
inhomogeneous modes} \hspace{\parindent} Beyond the tree level the
distribution function (\ref{2.25}) -- the diagonal element of the
reduced density matrix (\ref{2.25}) -- is no longer given by just
the square of the amplitude of the wavefunction. Now the
preexponential factor starts playing important role and, in
addition, nontrivial factor is being contributed due to the
integration over microscopic modes $f^T$. For no-boundary and
tunneling states these new contributions essentially modify the
tree-level algorithm. Interestingly, the expressions (\ref{i0})
and (\ref{i00}) get replaced by
        \begin{eqnarray}
        \rho_{\rm NB,T}(\varphi)\sim\exp[\mp I(\varphi)-
        \mbox{\boldmath$\SGamma$}(\varphi)],          \label{0.1}
        \end{eqnarray}
where the classical Euclidean action $I(\varphi)$ on the
quasi-DeSitter instanton with the inflaton field $\varphi$ -- the
4-dimensional sphere of the radius inverse to the Hubble constant
$H(\varphi)$ -- is amended by the loop effective action
$\mbox{\boldmath$\SGamma$}(\varphi)$ calculated on the same
instanton \cite{norm,tunnel,BarvU,tvsnb}. This action begins with
the one-loop functional determinant of the inverse propagator of
the full set of quantum fields $\phi$
        \begin{eqnarray}
        \mbox{\boldmath$\SGamma$}=
        \frac 12\,{\rm Tr}\,{\rm ln}\,\frac{\delta^2
        {\mbox{\boldmath $I$}}[\,\phi\,]}
        {\delta\phi\,\,\delta\phi}+...\,.       \label{IIII}
        \end{eqnarray}

The one-loop contribution can qualitatively change predictions of
the tree-level theory due to the dominant part of the effective
action induced by the anomalous scaling behaviour. On the
instanton of small size $1/H(\varphi)$ this asymptotic behaviour
looks like
        \begin{eqnarray}
        \mbox{\boldmath$\SGamma$}(\varphi)\sim Z\ln H(\varphi),
        \end{eqnarray}
where $Z$ is the total anomalous scaling of all quantum fields in
the model. This quantity is directly related to the conformal
anomaly of the theory integrated over the instanton volume. For
heavy fields it is dominated by the sum of terms quartic in their
mass parameters.

In the non-minimal inflaton model of \cite{qsi,qcr} the inflaton
potential is taken to contain the quartic term
       \begin{eqnarray}
        V(\varphi)=\frac{1}{2}m^{2}\varphi^{2}
        +\frac{\lambda}{4}\varphi^{4}.
        \end{eqnarray}
Also, it is natural to assume that this model contains generic GUT
sector of Higgs $\chi$, vector gauge $A_\mu$ and spinor fields
$\psi$ coupled to the inflaton via the interaction term
        \begin{eqnarray}
        {\mbox{\boldmath $L$}}_{\rm int}
        =\sum_{\chi}\frac{\lambda_{\chi}}4
        \chi^2\varphi^2
        +\sum_{A}\frac12 g_{A}^2A_{\mu}^2\varphi^2+
        \sum_{\psi}f_{\psi}\varphi\bar\psi\psi
        +{\rm derivative\,\,coupling}.             \label{0.3}
        \end{eqnarray}
For such a model the parameter $Z$ can be very big, because of the
Higgs effect generating large masses of all the particles directly
coupled to the inflaton. Due to this effect the anomalous scaling
(dominated by terms quartic in particle masses) is quadratic in
$|\xi|$,
        \begin{eqnarray}
        &&Z=6\frac{|\xi|^2}\lambda\mbox{\boldmath$A$},    \label{Z}\\
        &&{\mbox{\boldmath $A$}} = \frac{1}{2\lambda}
        \Big(\sum_{\chi} \lambda_{\chi}^{2}
        + 16 \sum_{A} g_{A}^{4} - 16
        \sum_{\psi} f_{\psi}^{4}\Big),               \label{A}
        \end{eqnarray}
with a particular coefficient ${\mbox{\boldmath $A$}}$ -- a
universal combination of the coupling constants above.

Thus, the probability peak in this model reduces to the extremum
of the function
        \begin{eqnarray}
        \ln\rho_{\rm NB,\,T}(\varphi)\simeq\mp I(\varphi)
        -3\frac{|\xi|^2}\lambda\mbox{\boldmath$A$}\,
        \ln\frac{\varphi^2}{\mu^2}.                    \label{1.1}
        \end{eqnarray}
in which the $\varphi$-dependent part of the classical instanton
(\ref{Inon}) action is very different from its analogue in the
minimal model (\ref{i000}). When expanded in powers of the slow
roll expansion parameter, $m_P^2/|\xi|\varphi^2\ll 1$, this action
equals
        \begin{eqnarray}
        &&I(\varphi)=-\frac{96\pi^2|\xi|^2}\lambda
        -\frac{24\pi(1+\delta)|\xi|}
        {\lambda}\frac{m_P^2}{\varphi^2}
        +O\,\left(\frac{m_P^4}{\varphi^4}\right),
        \label{1.2} \\
        &&\delta\equiv
        -\frac{8\pi\,|\xi|\,m^2}{\lambda\,m_P^2}.
        \label{delta}
        \end{eqnarray}

Thus the analysis of probability maxima does not reduce to
considering the extrema of the potential. Rather, at the
probability maximum the contribution of the instanton action gets
balanced by the anomalous scaling term, provided the signs of
$(1+\delta)$ and $\mbox{\boldmath$A$}$ are properly correlated
with the $(\mp)$ signs of the no-boundary (tunneling) proposals.
As a result the probability peak exists with parameters -- mean
values of the inflaton and Hubble constants and relative width
        \begin{eqnarray}
        &&\varphi_I^2= m_{P}^2
        \frac{8\pi|1+\delta|}{|\xi|
        {\mbox{\boldmath$A$}}},             \label{1.3}\\
        &&H^2(\varphi_I)=
        m_{P}^2\frac{\lambda}{|\xi|^2}
        \frac{2\pi|1+\delta|}
        {3{\mbox{\boldmath $A$}}},          \label{1.3a}\\
        &&\frac{\Delta\varphi}{\varphi_I}\sim
        \frac{\Delta H}{H}\sim
        \frac 1{\sqrt{12{\mbox{\boldmath $A$}}}}
        \frac{\sqrt{\lambda}}{|\xi|},              \label{1.4}
        \end{eqnarray}
which are strongly suppresed by a small ratio
$\sqrt{\lambda}/|\xi|$ known from the COBE normalization for
$\Delta T/T\sim 10^{-5}$ \cite{COBE,Relict}(because the CMBR
anisotropy in this model is proportional to this ratio
\cite{nonmin1,nonmin2}). This GUT scale peak gives rise to the
inflationary epoch with the e-folding number
        \begin{eqnarray}
        N\simeq \frac{48\pi^2}{\mbox{\boldmath $A$}}   \label{1.5}
        \end{eqnarray}
only for $1+\delta>0$ and, therefore, only for the {\it tunneling}
quantum state (plus sign in (\ref{1.1})). Comparison with $N\geq
60$ necessary for $\Omega>1$ immediately yields the bound on
$\mbox{\boldmath $A$}$ \cite{efeq},
        \begin{eqnarray}
        \mbox{\boldmath $A$}\sim 5.5,               \label{bound1}
        \end{eqnarray}
which can be regarded as a selection criterion for particle
physics models \cite{qsi,qcr}. These conclusions on the nature of
the inflation dynamics from the initial probability peak remain
true also at the quantum level -- with the effective equations
replacing the classical equations of motion \cite{efeq}.

\subsection{Open inflation without anthropic principle}
\hspace{\parindent} It is interesting that the synthesis of the
Hawking-Turok paradigm with the model of a strong non-minimal
coupling and a proper account of loop corrections gives a two-fold
result: it does not only generate a low-energy mechanism of
creation of the open Universe with a resonable value of $\Omega$
(not very close to zero or one) but also allows one to abandon the
anthropic principle inherent in the original model of
\cite{HawkTur}. This goes as follows \cite{openrev}.

The tree-level Euclidean action of the Hawking-Turok instanton in
the non-minimal model can be approximately calculated as an
expansion of the slow-roll parameter $m_P^2/|\xi|\varphi^2$,
similarly to the analogous expansion for the Hartle-Hawking
instanton (\ref{1.2}). In contrast with (\ref{1.2}), however, this
expansion has to be performed up to the second order in
$m_P^2/|\xi|\varphi^2$ inclusive (first order approximation turns
out to be insufficient as shown in \cite{openrev}) and it reads
        \begin{eqnarray}
        &&I_{HT}(\varphi)=-\frac{96\pi^2|\xi|^2}\lambda
        -\frac{2\pi(1+\delta)}\lambda \frac{m_P^2}{\varphi^2}
        +2\frac{(1+\delta)^2}\lambda
        \left(\frac{m_P^2}{\varphi^2}\right)^2
        \ln\left(\frac{6\pi|\xi|\varphi^2}
        {m_P^2(1+\delta)\kappa}\right)\nonumber\\
        &&\qquad\qquad\qquad\qquad\qquad\qquad\qquad
        +O\,\left(\frac{m_P^6}{|\xi|\varphi^6}\right),
        \label{3.16}
        \end{eqnarray}
where $\kappa$ absorbs the combination of numerical parameters
$\ln\kappa\equiv 11/3-\ln 4 - 3(1+2\delta)^2/4(1+\delta)^2$ (apart
from the negligible dependence of $\kappa$ on $\delta$,
$\kappa\sim 4.6$). Due to big $|\xi|$ it contains a large but
slowly varying (in $\varphi$) logarithmic term with positive
coefficient. It arises as a rather nontrivial contribution of the
singular point of the Hawking-Turok instanton and turns out to be
crucial for the mechanism of low-energy origin of open inflation.

With this Euclidean action the Hawking-Turok distribution function
(in both no-boundary and tunneling incarnations) have nontrivial
probability peaks already in the tree-level approximation, but
they correspond to the values of $\Omega$ too close either to one
or to zero. With the inclusion of the one-loop effective action
the situation qualitatively changes, just like in the closed case.
It turns out that the new probability peak arises. For the {\em
no-boundary} proposal, it has the following parameters -- the mean
value $\varphi_I$, Hubble constant $H_I$ and quantum dispersion:
        \begin{eqnarray}
        &&\varphi_I^2\simeq \frac{m_P^2}{|\xi|}\,
        \left(\frac2{3\mbox{\boldmath$A$}}
        \ln\frac{24\pi^2}
        {\kappa^2 e\,\mbox{\boldmath$A$}}
        \right)^{1/2},                                   \label{5.2}\\
        &&H_I^2\simeq m^2_P\,\frac\lambda{|\xi|^2}\,
        \frac1{12}
        \left(\frac2{3\mbox{\boldmath$A$}}
        \ln\frac{24\pi^2}
        {\kappa^2 e\,\mbox{\boldmath$A$}}\right)^{1/2},              \\
        &&\frac{\Delta\varphi}{\varphi_I}
        \sim\frac{\Delta H}{H_I}\sim
        \frac{\kappa^2\sqrt{6\mbox{\boldmath$A$}}}{72\pi^2}
        \,\frac{\sqrt\lambda}{|\xi|}.                      \label{5.3}
        \end{eqnarray}
Similarly to the closed model, these parameters are suppressed
relative to the Planck scale by a small dimensionless ratio
$\sqrt{\lambda}/|\xi|$ known from the COBE normalization. As
regards the e-folding number $N$, it is given for this peak
entirely in terms of the same universal combination of coupling
constants (\ref{A})
        \begin{eqnarray}
        N\simeq\left(\frac{24\pi^2}
        {\mbox{\boldmath$A$}}
        \ln\frac{24\pi^2}
        {\kappa^2 e\,\mbox{\boldmath$A$}}
        \right)^{1/2}.   \label{5.5}
        \end{eqnarray}
Comparison of this result with the e-folding number, $N\sim 60$,
necessary for generating the observable density $\Omega$,
$0<\Omega<1$, not very close to one or zero, immediately gives the
bound on $\mbox{\boldmath$A$}$
        \begin{eqnarray}
        \mbox{\boldmath$A$}\sim\frac{48\pi^2}{N^2}
        \ln\frac{N}{\kappa e}\sim 0.3.            \label{5.6}
        \end{eqnarray}

A similar analysis for the case of the {\em tunneling}
distribution function shows that its probability maximum
corresponds to the e-folding number, $N\simeq\kappa\sqrt e\sim 8$,
and the density parameter $\Omega\sim e^{-100}$ which are far too
small to describe the observable Universe. This leaves us with the
only candidate for the initial conditions of inflation
(\ref{5.2})-(\ref{5.3}) generated by the no-boundary Hawking-Turok
wavefunction of the open Universe.

\section{Conclusions}
\hspace{\parindent} Thus, despite conceptual and technical
problems, modern quantum cosmology represents viable theory
capable of predictions and fundamental (quantum gravitational)
justification of the origin of the inflationary Universe. It seems
to be consistently describing the low-energy phenomenon of such
origin which matches with the main observable cosmological
parameters -- CMBR anisotropy, admissible values of the density
parameter $\Omega$, restrictions on the duration of inflationary
epoch, etc. It also belongs the GUT energy scale which is much
below the Planckian one and, thus, can be justified within the
field-theoretical loop expansion. Unfortunately, these
phenomenologically attractive results do not resolve the
controversy between the no-boundary and tunneling cosmological
states. The tunneling state represents a viable scheme for a
closed Universe, while the tunneling one -- for an open model
within the Hawking-Turok prescription. Both states has the right
for existence and, apparently, wait for the proper moment when
they will be naturally included in some unifying framework, like
third quantization or theory of baby universes -- bright ideas
existing now, however, only at a very speculative level
\cite{bigfix}.

From the phenomenological viewpoint, quantum cosmology is expected
to be invoked for the explanation of such recently observed
phenomena like the acceleration of the Universe \cite{Lambda}. The
first attempt to use it as an alternative to quintessence
\cite{quint} has unfortunately failed \cite{efeqmy}.

In general, at the turn of new millennium the ideas of quantum
cosmology are anticipated to enrich other fields and to be unified
with other concepts of cosmological high energy physics and theory
of multi-dimensional compactifications induced by superstring
theory. For example, the origin of big non-minimal curvature
coupling of the inflaton in the model of the above type might be
accounted for within the Randall-Sundrum two-branes model
\cite{RanSun}, which as is known \cite{Gar} induces nonminimal
coupling in the effective 4-dimensional action. On the other hand,
the Hamilton-Jacobi equation method intensively used in context of
AdS/CFT correspondence and D-brane physics \cite{HJ} has its
origin in canonical quantum gravity and its cosmological
implications. It is good to see how old problems are finally
springing a leak and new revelations are expecting us.

\section*{Acknowledgements}
\hspace{\parindent} This work was supported by the Russian
Foundation for Basic Research under the grant No 99-02-16122, by
the grant of support of leading scientific schools No 00-15-96699
and also by the Russian Research program ``Cosmomicrophysics''.

\end{document}